\documentclass[prb,preprint]{revtex4}
\usepackage{graphicx}
\makeatletter
\usepackage{epsfig}
\def\@dotsep{4.5}
\usepackage{dcolumn}
\usepackage{amsmath}
\makeatother

\usepackage{graphicx}
\usepackage{bm}
\usepackage{amsfonts}
\usepackage{amsbsy}
\usepackage{amssymb}
\usepackage[mathscr]{eucal}

\newcommand{\beq}{\begin{equation}}
\newcommand{\eeq}{\end{equation}}
\newcommand{\ba}{\begin{array}}
\newcommand{\ea}{\end{array}}
\newcommand{\bea}{\begin{eqnarray}}
\newcommand{\eea}{\end{eqnarray}}
\newcommand{\bseq}{\begin{subequations}}
\newcommand{\eseq}{\end{subequations}}

\begin{document}

\title{The electron-phonon coupling strength at metal surfaces directly determined from the Helium atom scattering Debye-Waller factor}

\author{{\bf J. R. Manson}$^{1,2}$, {\bf G. Benedek}$^{2,3}$ and {\bf Salvador Miret-Art{\'e}s}$^{2,4}$}

\affiliation{
$^1$ Department of Physics and Astronomy, Clemson University, Clemson,
South Carolina 29634, U.S.A.\\
$^2$ Donostia International Physics Center (DIPC), Paseo Manuel de Lardiz{{a}}bal, 4
20018 Donostia-San Sebastian, Spain\\
$^3$ Dipartimento di Scienza dei Materiali, Universit{\`a} di Milano-Bicocca, Via Cozzi 53, 20125 Milano, Italy\\
$^4$ Instituto de F\'isica Fundamental, Consejo Superior de Investigaciones Cient\'ificas, Serrano 123, 28006 Madrid, Spain
}

\date{\today}

\begin{abstract}
A new quantum-theoretical derivation of the elastic and inelastic scattering probability of He atoms from a metal surface, where the energy and momentum exchange with the phonon gas can only occur through the mediation of the surface free-electron density, shows that the Debye-Waller exponent is directly proportional to the electron-phonon mass coupling constant $\lambda$. The comparison between the values of $\lambda$ extracted from existing data on the Debye-Waller factor for various metal surfaces and the $\lambda$ values known from literature indicates a substantial agreement, which opens the possibility of directly extracting  the electron-phonon coupling strength in quasi-2D conducting systems from the temperature or incident energy dependence of the elastic Helium atom scattering intensities.
\end{abstract}

\maketitle


\newpage

An atom at thermal energy scattered by a metal surface can exchange energy with the phonon gas of the solid through  oscillations of the electron density produced at the surface by the vibrational displacements of the atomic cores. Thus the inelastic atom scattering intensity when a phonon is created or annihilated has been shown to be approximately proportional to the electron-phonon (e-p) coupling constant (mass correction factor) $\lambda_{{\bf Q}, \nu}$  for that specific phonon mode of branch index $\nu$ and parallel wavevector ${\bf Q}$.
This was recently demonstrated for ultrathin lead films~\cite{Skl,Benedek-14} and the Bi(111) surface,~\cite{Tam-Benedek-13} enabling the so-called $mode-\lambda$ spectroscopy for the determination of individual phonon contributions to the mass correction factor.
The average of the coupling strength over the phonon spectrum defines the coupling constant $\lambda$
as $\lambda =  \langle \lambda_{{\bf Q}, \nu} \rangle$.~\cite{Allen,Grimvall}

In this letter we show that the application of  standard approximations of electron-phonon coupling theory for metals to the distorted wave Born approximation (DWBA) for atom-surface scattering leads to expressions which relate the elastic and inelastic scattering intensities, as well as the Debye-Waller (DW) factor, to the  mass correction factor of superconductivity theory.  This treatment, besides reproducing the previous result that the intensities for single-phonon inelastic peaks in the scattered spectra are proportional to the respective phonon mode components $\lambda_{{\bf Q}, \nu}$, leads to a new and useful formulation for the Debye-Waller factor $\exp\{-2W(T_S)\}$. Since the intensities of elastic  diffraction peaks are proportional to the Debye-Waller factor, their logarithmic temperature dependence at sufficiently large absolute temperature $T_S$  is a linear function of $T_S$ with a slope approximately proportional to the coupling constant $\lambda$.
Not only does this work demonstrate that $\lambda$ and its mode components can be measured in molecule-surface collision experiments, it also shows how phonon-induced modulation to the molecule-surface interaction potential is related to the surface electron coupling strength at the surface of metals.  Knowledge of the interaction potential is important for understanding adsorption, chemisorption or chemical reactions on metal surfaces.~\cite{Kroes-15,Wodtke-15}

In the distorted wave formalism employed here the important static parts of the potential such as the van der Waals  attraction with its associated adsorption well, and also the overall repulsive part, are contained in the distorted wave functions of the scattering atom.
The colliding He atom is repelled by the
surface electron density $n({\bf r})$.  In the absence of lattice vibrations
$n({\bf r})  ~=~ \sum_{{\bf K}, n}  \left|
\psi_{{\bf K}, n}({\bf r})  \right|^2$ where
$\psi_{{\bf K}, n}({\bf r}) = \exp\{ i {\bf K  \cdot R} \} \varphi_{{\bf K}, n}(z) $
are the electron wavefunctions of the occupied states which contribute to the surface density at the comparatively large distance ($\approx$3~\AA) from the first surface atomic plane where He atoms are repelled (classical turning point).  In practice $n({\bf r})$ receives the largest contribution from free electrons at the Fermi level, which allows for the factorization of wavefunctions into a simple plane wave of parallel wave vector ${\bf K}$ and band index $n$, and the wavefunction $ \varphi_{{\bf K}, n}(z)$
for the motion normal to the surface, with ${\bf r} = ({\bf R}, z)$.  Summations over $\{ {\bf K}, n \}$
implicitly include a factor of 2 for spin.
Multiplying this electron density by the Ebsjerg-N{\o}rskov constant $A_N$ gives the repulsive part of the distorting potential extending outside of the terminal surface layer of  core atoms.~\cite{Norskov,Senet,Zaremba}
The interaction potential $\mathcal{V} ({\bf r})$, which is regarded as a perturbation on the distorting potential, is then proportional to the variation in the electronic density caused by electron-phonon coupling
\begin{eqnarray} \label{L2}
\mathcal{V} ({\bf r}) =
A_N~ \delta n({\bf r})
=A_N ~\sum_{{\bf K}, n}  \left\{ \left| \tilde{\psi}_{{\bf K}, n} \right|^2 -
\Bigl| {\psi}_{{\bf K}, n} \Bigr|^2 \right\}
~,
\end {eqnarray}
where $ \tilde{\psi}_{{\bf K}, n}$
is the wave function with electron-phonon coupling to the cores and is related to
 $ {\psi}_{{\bf K}, n}$
through ordinary second order perturbation theory via the electron matrix elements
\begin{eqnarray} \label{L3}
\tilde{\psi}_{{\bf K}, n} = {\psi}_{{\bf K}, n} +  \sum^{~~~~~~\prime}_{{\bf K}^\prime, n^\prime}
{\psi}_{{\bf K}^\prime, n^\prime}
 \frac{ \left( {\psi}_{{\bf K}^\prime, n^\prime} \left| V^{el}  \right| {\psi}_{{\bf K}, n} \right)
}
{E^{el}_{{\bf K}, n}-E^{el}_{{\bf K}^\prime, n^\prime}}
~,
\end {eqnarray}
where the prime symbol  indicates that the state $\{ {\bf K}, n \}$ is excluded from the sum.

The electron-phonon interaction comes about through the electronic potential $ V^{el}$
written as a sum over the pairwise pseudopotentials $v^{el}({\bf r} - {\bf r}_\ell -{\bf u}_\ell(t) )$
between the electron at position ${\bf r}$ and the ions at the instantaneous positions
$  {\bf r}_\ell +{\bf u}_\ell(t) $, with ${\bf u}_\ell(t) $ being their vibrational displacements and $\ell$
a 3D integer index labeling the lattice sites.

In the following we shall use formal scattering theory with initial $(i)$ and final $(f)$ distorted states
$\exp\{{i {\bf K}_{i,f}  \cdot {\bf R} }\}~ \chi^{}_{k_{iz},k_{fz}}(z) ~ | n_{i,f} \rangle$
of the He-surface system, where $\chi^{}_{k_{iz},k_{fz}}(z) $ are the He atom initial and final distorted states for motion normal to the surface, and $| n_{i,f} \rangle$ are the many-body phonon states of the target crystal before and after scattering.  The transition rate for purely elastic diffraction is found to be
\begin{eqnarray} \label{L4}
\nonumber
w_{DWBA}^{(0)}({\bf k}_f , {\bf k}_i) =
 \frac{8 \pi  A_N^2}{\hbar a_c^2} \sum_{{\bf G}}~e^{-2W^{eff}( {\bf k}_f, {\bf k}_i) }
 \\ \nonumber
  \sum_{{\bf K}, n}~\left|\Re~\sum^{~~~~\prime}_{n^\prime}
  \frac{ \left( \chi^{}_{k_{iz}}(z)  \left| \varphi^*_{{\bf K}, n}(z)    \varphi_{{\bf K}+{\bf G}, n^\prime}(z)         \right| \chi^{}_{k_{fz}}(z)  \right) }
  {E^{el}_{{\bf K}, n}-E^{el}_{{\bf K}+{\bf G}, n^\prime}}
   \right.
  \\ \nonumber \times   \Biggl.
 ~\sum_{j} \left(\varphi_{{\bf K}+{\bf G}, n^\prime}(z)  \left| v^{el,eff}_{{\bf G}}(T_S, z-z_{j}) \right| \varphi_{{\bf K}, n}(z) \right)
 \Biggr|^2
 \\  \times
 \delta_{{\bf K}_f-{\bf K}_i, ~ {\bf G}} ~\delta( E_f-E_i)
 \, ,  ~~~~~
\end {eqnarray}
where $ \Re$ signifies the real part, $j$ is an index labeling the planes parallel to the surface and $a_c$ is the area of the unit cell.

As appears from Eq.~(\ref{L4}) the overlap integral  matrix element
$ ( \chi^{}_{k_{iz}}(z)  \left| \varphi^*_{{\bf K}, n}(z)
 \varphi_{{\bf K}+\Delta {\bf K}, n^\prime}(z)         \right| \chi^{}_{k_{fz}}(z)  )
 \equiv I_{F}( \Delta{\bf K}) $
is taken over a non-diagonal element of the electron density operator, which acts as the effective scattering potential of the He-atom distorted waves.  Contributions to this overlap integral come almost entirely from electrons near the Fermi energy, as indicated by the subscript $F$.
Outside the surface and in front of the terminating layer of cores the electron wave functions decrease roughly exponentially with a decay constant dictated by the work function.  The projectile wave functions are even more strongly decaying in the opposite direction into the surface electron density.  The overlap between these two opposing behaviors defines the thin 2-D surface region where the interaction takes place, and this essentially contains the locus of classical turning points.
Eq.~(\ref{L4}) is associated with an effective temperature-dependent Debye-Waller factor $ \exp\{-2W^{eff}( {\bf k}_f, {\bf k}_i) \}$
which accounts for the modulation of the surface charge density induced by the ion thermal fluctuations.
The remaining matrix element accounts for the scattering of the virtual electron from the lattice ion potential mediated on the thermal oscillations.  Thus also the ion pseudopotential is mediated on thermal ion fluctuations at the surface temperature $T_S$ and its 2D Fourier component is indicated in
Eq.~(\ref{L4}) as $  v^{el,eff}_{\Delta {\bf K}}(T_S, z-z_{j})$.
Eq.~(\ref{L4}) shows that the electron-phonon interaction induces a static corrugation in the electronic density that  produces diffraction peaks when the parallel momentum transfer
$\Delta {\bf K} = {\bf K}_f-{\bf K}_i$ is a surface reciprocal lattice vector ${\bf G}$.

The one-phonon transition rate allows us to
identify the effective Debye-Waller factor
$ \exp\{-2W^{eff}( {\bf k}_f, {\bf k}_i) \}$
that appears as a multiplicative factor in all scattered intensities, and its argument
is given by
\begin{eqnarray} \label{L5}
2 W^{eff}({\bf k}_f, ~{\bf k}_i)
  =
4
\sum_{{\bf Q}, \nu}   \frac{ \hbar }{ N M \omega_{{\bf Q},\nu}}~\sum_{{\bf K}, n}
\\ \nonumber \times
\left| \sum_{\alpha = 1}^3  \Delta k_\alpha
P_\alpha(T_S;{\bf Q}, {\bf K},\nu, n)  \right|^2
\Bigl[ n_{BE}(\omega_{{\bf Q},\nu}) + \frac{1}{2}\Bigr]
,
\end {eqnarray}
where $\hbar \Delta k_\alpha  = \hbar ({\bf k}_f-{\bf k}_i)_\alpha$ are cartesian components of the momentum transfer vector and $N$ is the number of atoms.
The electron-phonon coupling, which relates the mean square displacement of the electron density at the point of impact to that of the atomic cores, is contained in the factors of
\begin{eqnarray} \label{L5a}
P_\alpha(T_S;{\bf Q}, {\bf K},\nu, n) =
\\ \nonumber
 \Re\sum_{j}~\sum^{~~~\prime}_{n^\prime}
 \frac{I_{F}(\Delta{\bf K}) ~ e_\alpha({\bf Q},~\nu)}
 {\left( \chi^{}_{k_{iz}}(z)\left|~ \hat{q}_\alpha~n^{eff}_{{\bf Q}}(T_S,z)\right|\chi^{}_{k_{fz}}(z) \right)}
  \\ \nonumber \times
 \frac{  \left( \varphi^*_{{\bf K}-{\bf Q}, n^\prime}(z)~ \left|
\hat{q}_\alpha ~ v^{el,eff}_{ {\bf Q}}(T_S, z-z_{j})~ \right|~ \varphi_{{\bf K}, n}(z) \right)}
 {E^{el}_{{\bf K}, n}-E^{el}_{{\bf K}-{\bf Q}, n^\prime}}
,
\end {eqnarray}
where $\hat{q}_\alpha$ is a cartesian component of the vector operator  $\hat{{\bf q}} = \{ {\bf Q}, id/d z\}$
and $e_\alpha({\bf Q},~\nu)$ is a component of the phonon polarization vector.
The matrix element of the repulsive potential is
$( \chi^{}_{k_{fz}}(z) |~ \hat{q}_{\alpha}~n^{eff}_{{\bf Q}}(T_S,z) | \chi^{}_{k_{iz}}(z) )$ and it
receives its contribution entirely from the narrow region of the classical turning points.

It is noteworthy that the Ebsjerg-N{\o}rskov constant $A_N$ does not appear in the D-W factor.
This is an important observation because it means that the effective Debye-Waller factor is based only on the very general principle that the part of the atom-surface potential that gives rise to energy transfer is proportional to the surface electron density, but does not depend on the actual value of the proportionality constant.

Notable also is the fact  that the dependence on initial and final projectile momenta is substantially more complex than the simple quadratic dependence on $\Delta {\bf k}={\bf k}_f-{\bf k}_i$ encountered in  standard DW treatments, e.g., as in neutron scattering.
This more complex momentum dependence, as well as the dependence on the attractive adsorption well in the interaction potential, is introduced through the presence of the distorted atomic wave functions.
The polarization vectors as well as the mass $M$ in Eqs.~(\ref{L5}) and~(\ref{L5a}) are those of the crystal cores.   It is the electron-phonon coupling, via the e-p matrix elements, that produces the actual effective vibrational mean-square displacement vectors experienced by the colliding projectile interacting with the electron gas in front of the surface.
As an added comment, the Debye-Waller exponent contains additional dependence on the temperature over and above  that of the Bose-Einstein functions appearing explicitly.   This additional temperature dependence is contained in the effective potentials
$ n^{eff}_{{\bf Q}}(T_S,z) $
and $ v^{el,eff}_{ {\bf Q}}(T_S, z-z_{j})$ appearing in Eq.~(\ref{L5a})
and arises directly from the electron contribution to the Debye-Waller factor.

Rather than continuing with complete results, which are discussed in detail elsewhere,~\cite{Authors} it is of interest to apply  approximations valid for metals and to exhibit the inelastic intensity and Debye-Waller factor in terms of the standard definitions of the  e-p coupling constant.   The restriction to the Fermi level  allows the standard Grimvall approximation (especially appropriate for single phonon transitions)
$E^{el}_{{\bf K}, n}-E^{el}_{{\bf K}-\Delta {\bf K}, n^\prime}= \hbar \omega_{\Delta {\bf K},~\nu} $ with $\hbar \omega_{\Delta {\bf K},~\nu}$ the phonon energy.~\cite{Grimvall}
The electron-phonon coupling matrix is usually written with the following definition:~\cite{Grimvall,Allen,Skl,Eliashberg}
\begin{eqnarray} \label{L6}
\nonumber
g_{n,{n^\prime}}({\bf K},\,\Delta{\bf K},\, \nu)  \equiv
 \sum_{j}  \left[  \frac{ \hbar }{2 N M \omega_{\Delta {\bf K},~\nu}} \right]^{1/2}
 {\bf e}(\Delta {\bf K},\nu)
\\
\cdot
  \left( \varphi^*_{{\bf K}-\Delta {\bf K}, n^\prime}(z) \left|
\hat{{\bf q}} ~ v^{el,eff}_{\Delta {\bf K}}(T_S, z-z_{j}) \right|~ \varphi_{{\bf K}, n}(z) \right),
~~~~
\end {eqnarray}
and the mode-specific components of $\lambda$ are~\cite{Allen}
\begin{eqnarray} \label{L6b}
\lambda_{ {\bf Q}, \nu} =\frac{2}{\mathcal{N}(E_F) [\hbar   \omega_{{\bf Q},\nu} ]^3}
\sum_{{\bf K}, n}
\Biggl| \sum^{}_{n^\prime}
g_{n,{n^\prime}}({\bf K},{\bf Q}, \nu)
\Biggr|^2
,
\end {eqnarray}
where the summation over electron wave vectors involves only states near the Fermi surface
and  $\mathcal{N}(E_F)$ is the density of electron states at the Fermi surface.
The definition for
$g_{n,{n^\prime}}({\bf K},\,\Delta{\bf K},\,\, \nu) $ and hence that for $\lambda_{ {\bf Q}, \nu} $, differs slightly from the standard definitions~\cite{Grimvall,Allen} in that an effective, temperature-dependent electron-phonon potential now appears in the matrix element in our treatment.  This arises from the
electronic
Debye-Waller considerations that we have introduced into this treatment.

\begin{figure}
\includegraphics[width=2.5in]{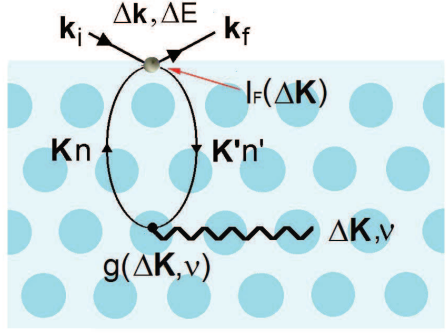}
\caption{An incident atom in a state   of wavevector ${\bf k}_i$  is inelastically scattered into a final state   of wavevector ${\bf k}_f$ by the overlap vertex $I_F(\Delta{\bf K})$ and creates a phonon of wavevector $\Delta{\bf K}$ and branch index $\nu$ via a virtual electron-hole pair of states   and    the electron-phonon vertex term $g$.  }
\label{Diagram}
\end{figure}
The inelastic He-atom scattering (HAS) process is illustrated by the diagram of Fig.~\ref{Diagram}, in which an incident projectile atom in a state   of wavevector ${\bf k}_i$ is inelastically scattered into a final state   of wavevector ${\bf k}_f$, eventually creating a phonon of wavevector ${\Delta \bf K}$ and branch index $\nu$ via  the mediation of a virtual electron-hole pair involving the electronic states     at the Fermi level of parallel wavevectors ${\bf K}$ and ${\bf K}^\prime$ and band indices $n$ and $n^\prime$. The non-diagonal electron density matrix element $I_F(\Delta {\bf K})$  acts as an effective scattering potential, whose matrix element between the initial and final atom states provides the upper vertex term, whereas the lower vertex term is expressed by the electron-phonon matrix element $g_{n,{n^\prime}}({\bf K},\,\Delta{\bf K},\, \nu) $. It is important to remark that the phonon can be generated near the surface or, as depicted in Fig.~\ref{Diagram}, at several atomic planes beneath the surface, the maximum depth being determined by the range of the e-p interaction (quantum sonar effect).~\cite{Skl,Benedek-14,Authors}

Combining the above  e-p approximations and definitions casts the single phonon inelastic transition rate into the form
\begin{eqnarray} \label{L7}
w_{{DWBA}}^{(1)}({\bf k}_f , {\bf k}_i) =
\frac{4 \pi A_N^2}{\hbar a^2_{c}}~
  \mathcal{N}(E_F)
  e^{-2W^{eff}( {\bf k}_f, {\bf k}_i)}
 \\ \nonumber \times
 ~\sum_{\nu}~  \hbar   \omega_{\Delta {\bf K},\nu}
 ~\left| I_{F}(\Delta {\bf K}) \right|^2
 ~ \lambda_{\Delta {\bf K}, \nu}
  \\ \nonumber \times ~
\left| n_{BE}(\omega_{\Delta {\bf K},\nu}) \right|
~\delta \bigl(E_f-E_i - \hbar   \omega_{\Delta {\bf K},\nu}  \bigr)
~.
\end {eqnarray}
This is similar to the new result of Sklyadneva {\em et al.}~\cite{Skl,Benedek-14}
showing that the probability of creating (or annihilating) a phonon mode $\{{\bf Q},\,\nu\}$ with frequency $\omega_{{\bf Q},\nu}$ is proportional to the respective mode dependent electron-phonon coupling constant $ \lambda_{{\bf Q}, \nu}$.  However, Eq.~(\ref{L7}) contains an effective DW factor, an important difference with respect to the previous work~\cite{Skl,Benedek-14}
where the one-phonon approximation was made before the thermal average over the phonon ensemble and no DW factor is found in that case.

It is of interest to examine the Debye-Waller exponent $2 W^{eff}$ under reasonable approximations applicable to elastic diffraction.  Since the Debye-Waller argument is expressed as a sum over all contributing phonon modes  for a given set of experimental initial incident beam and final detector parameters, it will of necessity be expressed as a weighted summation over the mode specific ${\lambda}_{{\bf Q}, \nu}$.  The simplest case, and the configuration that is most often measured, is the specular diffraction peak in which the parallel momentum exchange vanishes and the total momentum transfer is $2 k_{iz}$ entirely in the direction normal to the surface.
In this case, by considering an exponential decay $\exp\{-2 \kappa z\}$ for the electron density at the He-atom turning point, with $ \kappa = \sqrt{2m^*_e \phi}/\hbar$, $\phi$ is the work function and $m^*_e$ is the electron effective mass, leads to
\begin{eqnarray} \label{L8}
2 W^{eff}({\bf k}_f, ~{\bf k}_i)  =
\frac{\mathcal{N}(E_F)}{3N} \frac{m}{m^*_e}  \frac{E_{iz}}{\phi}
\sum_{{\bf Q}, \nu}  \hbar \omega_{{\bf Q},\nu}   {\lambda}_{{\bf Q}, \nu}
\nonumber
\\  \times
  \left|\frac{I_F({\bf Q})}{
  \left( \chi^{}_{k_{iz}}(z)\left| n^{eff}_{{\bf Q}}(T_S,z)\right|\chi^{}_{k_{fz}}(z) \right)
  } \right|^2
 \Bigl[ n_{BE}(\omega_{{\bf Q},\nu}) + \frac{1}{2}\Bigr]
,
\end {eqnarray}
where $m$ is the He atom mass and $E_{iz} = \hbar^2 k_{iz}^2/2m$ the normal part of the He atom incident energy.
Eq.~(\ref{L8}) shows  the Debye-Waller exponent  expressed explicitly as a weighted summation over
${\lambda}_{{\bf Q}, \nu}$.  The weighting coefficients are quantities that can be readily evaluated, i.e.,  the squared ratio of the overlap integral and the distorted wave matrix elements of the repulsive potential,  and in many cases this ratio is nearly unity, for example in the case of weakly corrugated metals.  Thus in the high temperature limit where $n_{BE} \rightarrow k_B T_S/\hbar \omega_{{\bf Q},\nu}$ the Debye-Waller exponent for specular diffraction is proportional to $\lambda$:
\begin{eqnarray} \label{L9}
2 W^{eff}({\bf k}_f, {\bf k}_i)  \cong
\mathcal{N}(E_F) \frac{m}{m^*_e}  \frac{E_{iz}}{\phi}
\lambda  k_B  T_S
\end {eqnarray}
Thus the coupling constant $\lambda$ can be directly obtained from the temperature dependence of the HAS specular intensity $\mbox{I}_{00}$ as
\begin{eqnarray} \label{L9a}
\lambda_{HAS}  \cong  -\frac{1}{\mathcal{N}(E_F)}
\frac{d \ln \mbox{I}_{00}}{k_B d T_S}   \frac{m^*_e}{m}  \frac {\phi}{E_{iz}}
\end {eqnarray}
provided the electronic density of states and effective mass at the Fermi level are known.  In general the tail of the electron density far away from the surface atomic plane, where the He atoms are reflected, receives a major contribution from the surface states at the Fermi level.  For a simple estimation the free electron expression
$ m^*_e / \mathcal{N}(E_F)=  \hbar^2 k_F^2/ 3 Z $ may be used, where $k_F$ is the Fermi wavevector and $Z$ is the number of free electrons per atom.

\begin{figure}
\includegraphics[width=3.0in]{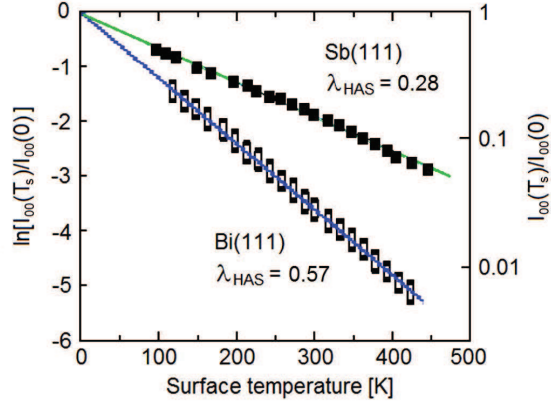}
\caption{Debye-Waller plots of the specular intensity vs. $T_S$ for He atom scattering from Sb(111) and Bi(111).  Data for Sb(111) from Ref.~\cite{Ernst-Sb-SurfSci617-13} and for Bi(111) from Ref.~\cite{Ernst-Bi-JPCM-12}. }
\label{FigDW}
\end{figure}

As an example we show in Fig.~\ref{FigDW} recent measurements of the logarithmic thermal attenuation (DW plots) of the HAS specular intensity from the Sb(111)\cite{Ernst-Sb-SurfSci617-13} and Bi(111)\cite{Ernst-Bi-JPCM-12} surfaces.  The corresponding values of $\lambda_{HAS}$ extracted from Eq.~(\ref{L9a}) under the free electron assumption are 0.28 and 0.57, respectively, and compare favorably with the surface values reported by Hofmann for Bi(111)\cite{d} and calculated ab-initio by Campi {\em et al.} for Sb(111) \cite{q}.
A list of these and other values for a selection of metal surfaces appears in Table~\ref{lambdatable}, and the agreements between the surface and known bulk values of $\lambda$
are quite reasonable.
\begin{table}
\caption{The mass enhancement factor $\lambda_{HAS}$ expressing the electron-phonon interaction as derived from the temperature dependence of the
HAS specular intensity (Eq.~(\ref{L9a}) with the free-electron assumption) for selected conducting surfaces and compared with values of $\lambda$ from other sources as cited.
In the column marked Pb(111) the experimental data from Ref.~[\cite{j}] and the value given for $\lambda_{HAS}$  are for seven monolayers of Pb on a Cu(111) substrate.}
\vspace{1cm}
\centering
\begin{tabular}{|c||c|c|c|c|c|}
  \hline      Surface &  Bi(111)  & Cu(110) & Pb(111) & Sb(111)  & W(001)1X1 \\
  \hline     $  {-d \ln \mbox{I}_{00}} / {dT_S} \left[ {10^{-3} }{\mbox{K}^{-1}} \right]  $  &  11.5\cite{Ernst-Bi-JPCM-12}  &1.7\cite{e} & 5.0\cite{j}  & 5.6\cite{Ernst-Sb-SurfSci617-13}
   & 4.1\cite{r} \\
  \hline     $ k_{iz}^2 \left[\mbox{\AA}^{-2}\right] $ & 16.79 \cite{Ernst-Bi-JPCM-12} & 6.20\cite{e} & 5.65\cite{j,lx}  & 22.8\cite{Ernst-Sb-SurfSci617-13}
   & 26.3\cite{r} \\
  \hline     $  \phi \left[ \mbox{eV} \right]$  &  4.23 \cite{b}   & 4.48 \cite{f} & 4.25\cite{k} & 4.56 \cite{o}
  & 4.32 \cite{s} \\
  \hline     $ k_F \left[ \mbox{\AA}^{-1} \right]$  &  0.72 \cite{c}   & 0.25 \cite{g} & 0.65\cite{Benedek-14} &  0.80 \cite{p}  & 1.09 \cite{t} \\
  \hline     $ \lambda_{HAS}$ &  0.57   & 0.15  & 0.76 & 0.28  & 0.31 \\
  \hline     $ \lambda$~(other sources) &  0.60 \cite{d}   & 0.17\cite{Jiang-Chulkov-14}  & 0.95\cite{Benedek-14} & 0.27 \cite{q}  & 0.28 \cite{u} \\
     &  0.45\cite{Ortizoga-14}   &  {0.23 \cite{i,ii}} & 0.7-0.9\cite{Zhang-05}   &  & \\ \hline
\end{tabular}
\vspace{1cm}
\label{lambdatable}
\end{table}

Some comments about the data presented in Table~\ref{lambdatable} are in order.  The experimental HAS data for Pb(111) taken from Refs.~[\cite{j}] and [\cite{lx}]  are for the specific case of seven monolayers of Pb on a Cu(111) substrate.  It is also evident from the list of measured values of $\lambda$ from other sources, particularly for the cases of Bi(111) and Cu(110), that there can be significant variations in reported values.
Finally, a value of $\lambda=1.3$ for Bi(111) has recently been reported.~\cite{Tam-Benedek-13}
This value, which was obtained by comparing energy-resolved inelastic HAS scattering spectra for Bi(111) with those of Pb(111), is larger than that reported here  obtained from Debye-Waller factor measurements on the specular diffraction peak, and also larger than the values reported from other sources.   However, inelastic HAS spectra sample only a limited number of phonon modes, namely only those modes that can be accessed along the He atom scan curve.  The scan curve, which results from energy and momentum conservation parallel to the surface, appears in a plot of energy transfer vs. $\Delta {\bf K}$ as a parabolic function whose shape is determined by the incident He atom beam energy and angles, and the angular position of the detector.
Only those phonons with both energy $\hbar \omega_{\Delta {\bf K}, \nu}$ and parallel momentum $\Delta {\bf K}$ lying on the scan curve are accessible in an inelastic HAS spectrum for a given set of experimental conditions.
When, as a trial estimate of $\lambda$ , the average over $\{{\bf Q}, \nu\}$ is restricted to a selected set of phonons, e.g., those sampled by HAS along a single scan curve,   trial values of $\lambda$ dispersed over a fairly large range may be found.   This is illustrated by the interesting example of Bi(111) vs. Pb(111) in Ref.~[\cite{Tam-Benedek-13}], where a ratio of $\lambda_{Pb}/\lambda_{Bi} = 1.35$
was obtained from similar HAS scan curves for each of the two metals. This is in contrast with the ratio of 0.75, about a factor 2 smaller, from the Debye-Waller factors reported in Table~\ref{lambdatable}, and means that under the kinematic conditions of Ref.~[\cite{Tam-Benedek-13}] the phonons sampled in Bi(111) make a larger contribution to e-p interaction than in the 7ML-Pb(111) film. The comparison is interesting because Pb is a bulk superconductor with transition temperature decreasing from bulk to ultrathin films, whereas bulk Bi is not a superconductor, but it becomes so in reduced dimensionality~\cite{Weitzel-91}, possibly because of some specific surface-localized phonons with  particularly strong e-p coupling. Thus, sampling  different segments of the phonon spectrum with inelastic HAS may help in pinpointing which phonons are actually important for electron pairing in low-dimensional superconductors.
The Debye-Waller factor, on the other hand, is not similarly limited by the scan curve, and hence is able to produce the true and correctly averaged $\lambda$ for a given surface.

This treatment, based on electron-phonon interaction theory, determines the corrugation and vibrational displacements of the electron gas in front of a surface at the locus of points where an incoming atom is reflected.  All aspects of the scattering are related to the electron-phonon coupling constants.  The single-phonon scattered intensity shows that the vibrational displacements of the electron density above the surface may be quite different from those of the core atoms, and even the polarizations for the same phonon mode may be in different directions.  However, the displacements of the electron density are expressed directly in terms of the masses and polarization vectors of the crystal cores.
Of particular importance, the single phonon scattering intensities are proportional to the mode-specific electron-phonon constants ${\lambda}_{{\bf Q}, \nu}$ and the argument of the Debye-Waller factor is to a good approximation proportional to the electron-phonon coupling constant $\lambda$.

Acknowledgment: one of us (GB) would  like to thank Prof. Marco Bernasconi for helpful discussions.
This work is partially supported by a  grant with Ref.
FIS2014-52172-C2-1-P from the Ministerio
de Economia y Competitividad (Spain).



\end{document}